\documentclass[prb,twocolumn,showpacs,preprintnumbers,amsmath]{revtex4}
\usepackage{dcolumn}
\usepackage{bm}
\usepackage{graphicx}
\usepackage{color}
\begin{document}
\title{Multi-gap superconductivity in single crystals of Ba$_{0.65}$Na$_{0.35}$Fe$_2$As$_2$: A calorimetric investigation}

\author{A. K. Pramanik}\email{a.k.pramanik@ifw-dresden.de, ashpramanik@gmail.com}\affiliation{Institute for Solid State Research, IFW Dresden, D-01171 Dresden, Germany}
\author{M. Abdel-Hafiez}\affiliation{Institute for Solid State Research, IFW Dresden, D-01171 Dresden, Germany}
\author{S. Aswartham}\affiliation{Institute for Solid State Research, IFW Dresden, D-01171 Dresden, Germany}
\author{A. U. B. Wolter}\affiliation{Institute for Solid State Research, IFW Dresden, D-01171 Dresden, Germany}
\author{S. Wurmehl}\affiliation{Institute for Solid State Research, IFW Dresden, D-01171 Dresden, Germany}
\author{V. Kataev}\affiliation{Institute for Solid State Research, IFW Dresden, D-01171 Dresden, Germany}
\author{B. B\"{u}chner}\affiliation{Institute for Solid State Research, IFW Dresden, D-01171 Dresden, Germany}

\date{\today}

\begin{abstract}
We investigate the electronic properties and the superconducting
gap characteristics of a single crystal of hole-doped 122
Fe-pnictide Ba$_{0.65}$Na$_{0.35}$Fe$_2$As$_2$ by means of
specific heat measurements. The specific heat exhibits a
pronounced anomaly around the superconducting transition
temperature $T_c$ = 29.4 K, and a small residual part at low
temperature. In a magnetic field of 90 kOe, the transition is
broadened and $T_c$ is lowered insignificantly by an amount $\sim$
1.5 K. We estimate a high electronic coefficient in the normal
state with a value 57.5 mJ mol$^{-1}$ K$^2$, being consistent with
hole-doped 122 compounds. The temperature-dependent
superconducting electronic specific heat cannot be described with
single-gap BCS theory under weak coupling approach. Instead, our
analysis implies a presence of two $s$-wave like gaps with
magnitudes $\Delta_1(0)/k_BT_c$ = 1.06 and $\Delta_2(0)/k_BT_c$ =
2.08 with their respective weights of 48\% and 52\%. While our
results have qualitative similarities with other hole-doped 122
materials, the gap's magnitude and their ratio are quite
different.
\end{abstract}


\maketitle
\section{Introduction}
The recent discovery of superconductivity (SC) in Fe-based
pnictides\cite{kamihara} has lead to wide research activities in
both experimental and theoretical frontiers of solid state
physics.\cite{paglione,johnston} This is primarily because
Fe-pnictides exhibit a high transition temperature ($T_c$), a
layered structure, and a proximity between magnetism and SC in its
phase diagram - a scenario reminiscent of cuprates. However,
Fe-pnictides are multiband metals where all five Fe-3$d$ orbitals
contribute to the electronic structure in the vicinity of the
Fermi surface (FS) having a stark contrast with cuprates which are
single band Mott-insulators. Detailed band structure
calculations, indeed, show that the FS in Fe-pnictides is
characterized by two electron-like cylinders around the $M$ point,
and two hole-like cylinders plus a 3D heavy hole-like pocket
around the $\Gamma$ point, thereafter implying it a possible
multiband superconductor.\cite{singh1111,mazin}

As for superconductors in general, the central issue remains to
understand the superconducting gap symmetry and the mechanism for
Cooper pairing, which are still under debate in case of
Fe-pnictides. With a weak electron-phonon coupling in this class
of materials,\cite{mazin,boeri,boeri-c} theoretical
calculations predict unconventional SC mediated by
antiferromagnetic (AFM) spin fluctuations, and an $s_{\pm}$ type
superconducting gap symmetry where the order parameter requires a
sign change between different sheets of the
FS.\cite{mazin,kuroki,chubu} Experimental findings
yet exhibit no consensus on the gap symmetry. For
instance, nearly isotropic two full gaps with different magnitudes
are evidenced in angle resolved photoemission spectroscopy (ARPES)
experiments for both electron- and hole-doped 122
Fe-pnictides.\cite{ding,evtu1,terashima,wray,evtu2} A similar
situation is observed in other studies, like, point contact
Andreev reflection spectroscopy (PCARS),\cite{samuely} and
penetration depth measurements.\cite{khasanov,williams,martin} On
the other hand, possible existence of nodes is revealed in the
temperature ($T$) dependence of penetration depth
and nuclear magnetic resonance (NMR) measurements for both the 122
and 1111 series.\cite{gordon,grafe} It can be mentioned that most
of these investigations (except NMR) are surface sensitive,
therefore, sample impurity or inhomogeneity at the surface may
cause such contradicting results.

In this situation, specific heat ($C$) rather
provides a key source of information regarding the bulk
thermodynamic properties, exploring the electronic- and gap-structure
in materials. In addition, it probes the system in
equilibrium and low energy state. Recent specific heat studies in
different families of Fe-pnictides have explored dissimilar gap
properties with a single to double gaps and even the presence of
nodes, and this variation appears related to the nature as well
as the level of
doping.\cite{mu1111,mu122,hardy,hardy1,kant,popovich,gofryk,fukazawa}

In this contribution we investigate the electronic properties and
superconducting gap characteristics in a hole-doped 122 compound,
Ba$_{0.65}$Na$_{0.35}$Fe$_2$As$_2$ ($T_c$ = 29.4 K), by means of
specific heat measurements. Although, such investigations have
been performed in great details for its K-doped analogues
Ba$_{1-x}$K$_{x}$Fe$_2$As$_2$, such studies are lacking in the
Na-doped 122 family. It is, however, necessary to scrutinize how
these properties are sensitive to different dopant species with
unlike sizes and chemistry, considering the fact that
the gap properties significantly modify with the
nature of the doping elements. To be precise, ARPES experiments
reveal that for hole-doped Ba$_{0.6}$K$_{0.4}$Fe$_2$As$_2$, the
average gap for the inner hole-like and two electron-like
cylinders is similar with a large value $\sim$ 12 meV, while the
outer hole cylinder is having a lower value $\sim$ 6
meV.\cite{ding} In contrast, for electron-doped
BaFe$_{1.85}$Co$_{0.15}$As$_{2}$, the study shows that the inner
hole pocket disappears and the average gaps of comparable sizes
(6.6 and 5 meV) are observed in the outer hole and two electron
cylinders.\cite{terashima} These observations underline that
detailed investigations are necessary in different compositional
materials for a generalized understanding of these issues.

Our results show a pronounced specific heat anomaly at $T_c$ where
the jump height is consistent with the $T_c$ value according to
recent results on Fe-pnictides. This transition is minimally
suppressed in magnetic fields of 90 kOe. Our estimated electronic
coefficient in the normal state is high in agreement with other
hole-doped 122 compounds. Our analysis further shows that the
superconducting electronic specific heat cannot be described with
the single-band weak-coupling BCS scheme, rather it implies the
presence of two $s$-wave like gaps with different magnitudes and
contributions. These results have qualitative
similarities with K-doped materials. However, the quantitative
difference in the gap ratio may indicate the different density of
states (DOS) in respective bands and the different interband
interaction in these materials, hence highlighting the element
specific role of the dopant.

\section{Experimental Details}
Single crystals of Ba$_{0.65}$Na$_{0.35}$Fe$_2$As$_2$ (BNFA) and
BaFe$_2$As$_2$ (BFA) used in the present study have been grown
using a self-flux method. The details of sample preparation and
characterization are described elsewhere.\cite{sai,sai1} The
parent compound, i.e., BaFe$_2$As$_2$ has been used to estimate
the lattice specific heat contribution. The crystals have been
characterized with x-ray diffraction (XRD) which implies the
absence of any chemical impurity phase within the experimental
accuracy. The mentioned chemical compositions of this material have been determined by energy
dispersive analysis of x-ray (EDAX) spectroscopy performed at
different places of the sample. For the sample BNFA, the Na variation
in the used piece is found within the instrumental error
limit. A recent study on polycrystalline Ba$_{1-x}$Na$_{x}$Fe$_2$As$_2$
shows an unstable crystallographic phase in Na-rich compositions where the material
is susceptible to chemical impurity phases owing to the large
mismatch in the size of Ba$^{2+}$ and Na$^{1+}$ ions.\cite{gil}
In view of this, our results regarding the crystal homogeneity are remarkable.
The BNFA crystal used in this study is about 2.47
$\times$ 1.98 $\times$ 0.13 mm$^3$. The magnetization ($M$) data
have been collected using a SQUID-VSM magnetometer made by Quantum
Design. The heat capacity is measured along the crystallographic $c$ axis
with a Physical Property
Measurement System (Quantum Design) using a thermal relaxation
technique down to 1.8 K and magnetic fields up to 90 kOe.

\begin{figure}
    \centering
        \includegraphics[width=8.5cm]{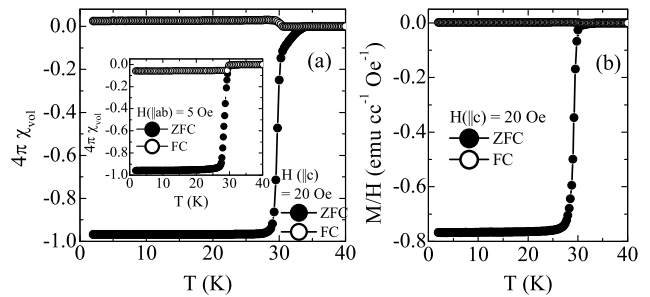}
    \caption{(a) The volume susceptibility $\chi_{vol}$ after demagnetization correction has been plotted as function of temperature. The $\chi_{vol}$ has been deduced from the dc magnetization measured with $H ||c$ = 20 Oe following ZFC and FC protocols for Ba$_{0.65}$Na$_{0.35}$Fe$_2$As$_2$. The inset shows the similarly deduced $\chi_{vol}$ with $H || ab$ plane ($H$ = 5 Oe). (b) The same data in main panel of (a) have been plotted without demagnetization correction.}
    \label{fig:Fig1}
\end{figure}

\section{Results and Discussions}
In Fig. 1a we present the temperature dependence of the volume
susceptibility ($\chi_{vol}$) measured following the
zero-field-cooled (ZFC) and field-cooled (FC) protocols for BNFA.
$\chi_{vol}$ has been deduced from the dc-magnetization data
measured in a field of 20 Oe applied parallel to the $c$ axis.
Care has been taken to correct the magnetization data for
demagnetization effect where the demagnetization factor has been
estimated from an ellipsoidal approximation based
on the dimensions of the crystal.\cite{osborn} The material
exhibits bulk SC which is evident from the diamagnetic signal in
the $M_{ZFC}$ curve at low temperatures. Although, the ZFC and FC
magnetization already start to bifurcate around 34 K, our material
shows a sharp superconducting transition (width $\sim$ 1.5 K)
around 29.7 K determined from the inflection point in
$dM_{ZFC}/dT$. However, similarly deduced $\chi_{vol}$ with $H || ab$ plane ($H$ = 5 Oe)
shows a clear bifurcation between the ZFC and FC magnetization data around 29.5 K,
as evident from the inset of Fig. 1a. This difference in onset of bifurcation between
our ZFC and FC magnetization data for fields applied along different crystallographic
directions will be studied in more detail in the future. Notably, $\chi_{vol}$ exhibits
an almost full diamagnetic shielding at low temperatures with fields parallel to
both $c$ axis and $ab$ plane. These results are in support of the good quality of our single crystal.
The fact, that $M_{FC} > 0$ for $H || c$ axis within the superconducting state seems to be an
artifact in the data, probably arising from flux trapping during
the FC process which is likely for this field geometry due to layered structure in this material.
Fig. 1b shows the same data presented in main panel of Fig. 1a without demagnetization correction, demonstrating
anomalies are not significantly evident. We would like to mention that the onset of
negative magnetization at temperatures higher than the sharp
transition in Fig. 1 with $H || c$ axis is not visible in the specific heat data
which exhibit a sharp jump around 29.4 K observed in the $M_{ZFC}$
curve (shown below).

The temperature dependence of the specific heat in the form $C/T$
vs $T$ is shown in Fig. 2 for BFA in 0 Oe and for BNFA in 0 and 90
kOe. For our further analysis, the electronic contribution to the
specific heat ($C_{el}$) is required for the material under study
(BNFA). Since BNFA is nonmagnetic, the subtraction of the lattice
specific heat ($C_{ph}$) from the total specific heat ($C_{tot}$)
will simply serve our purpose. Conventionally, $C_{ph}$ is
estimated by suppressing the superconducting transition in high
magnetic fields. However, the upper critical field ($H_{c2}$) is
significantly high in this class of superconductors. Thus, we have
estimated $C_{ph}$ from its parent compound BFA, which is not
superconducting throughout the temperature range, as evident from
Fig. 2 where $C/T$ does not exhibit any anomalous
behavior as function of temperature. On cooling from room
temperature, BFA exhibits a long-range magnetic order of AFM-type
paired with spin density wave (SDW) formation around 140 K,\cite{rotter,huang}
implying a likely magnetic contribution to its specific
heat. In fact, our specific heat data show a
sharp peak around this AFM-SDW transition in BFA (not shown).
However, a recent neutron scattering measurement has revealed that
the energy gap for low-energy spin-wave excitations in the
magnetically ordered state is about 9.8 meV ($\equiv$ 114 K) for this
material.\cite{matan} Therefore, magnetic contributions to the
specific heat will be negligible in the range of our working
temperatures ($<$ 35 K), and the specific heat can be assumed to
consist of electronic and lattice contributions only.

\begin{figure}
    \centering
        \includegraphics[width=8cm]{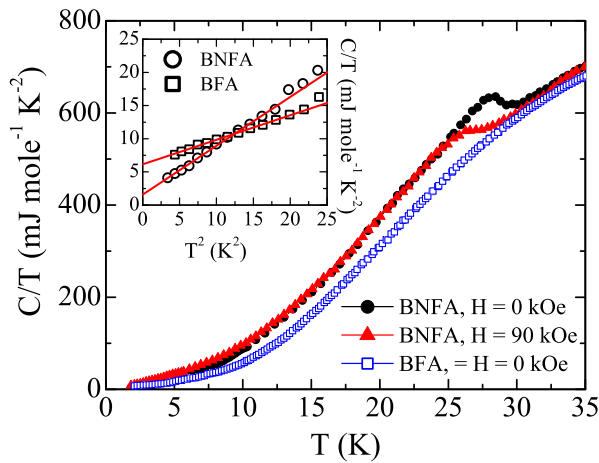}
    \caption{(Color online) Temperature dependence of the specific heat $C/T$ measured in 0 and 90 kOe for Ba$_{0.65}$Na$_{0.35}$Fe$_2$As$_2$ and BaFe$_2$As$_2$. The inset shows the plot $C/T$ vs $T^2$. The straight lines represent linear fits to $C/T$ = $\gamma$ + $\beta T^2$ (see text).}
    \label{fig:Fig2}
\end{figure}

\begin{table*}
\caption{\label{tab:table 1} The superconducting transition
temperature $T_c$, the jump height of the electronic specific heat
$\Delta C_{el}/T_c$, the normal state electronic specific heat
coefficient $\gamma_n$, and the superconducting gap properties
$\alpha_i$ and $\gamma_i/\gamma_n$ extracted from specific heat
measurements for Ba$_{0.65}$Na$_{0.35}$Fe$_2$As$_2$ along with
other hole- and electron-doped 122 Fe-pnictides. The $\alpha_i$
and $\gamma_i/\gamma_n$ represent the zero temperature gap ratio
and its weight in the $i$-th band, respectively.}
\begin{ruledtabular}
\begin{tabular}{cccccc}
Compounds &Ref. &$T_c$ &$\Delta C_{el}/T_c$ &$\gamma_n$  &$\alpha_i=\Delta_i(0)/k_BT_c$, $\gamma_i/\gamma_n$\\
 & &($K$) &(mJ mol$^{-1}$ K$^{-2}$) &(mJ mol$^{-1}$ K$^{-2}$) &\\
\hline
Ba$_{0.65}$Na$_{0.35}$Fe$_{2}$As$_{2}$\footnotemark[1] &This work &29.4 &72.5 &57.5 &$\alpha_1$=1.06, $\gamma_1/\gamma_n$= 0.48\\
& & & & &$\alpha_2$=2.08, $\gamma_2/\gamma_n$= 0.52\\
\hline
Ba$_{0.6}$K$_{0.4}$Fe$_{2}$As$_{2}$\footnotemark[1] &\onlinecite{mu122} &36.5 &98.1 &63.3 &$\alpha_1$=1.945, $\gamma_1/\gamma_n$= 1\\
\hline
Ba$_{0.6}$K$_{0.4}$Fe$_{2}$As$_{2}$\footnotemark[2] &\onlinecite{kant} &37.3 &$\sim$100 &49 &$\alpha_1$=2.07, $\gamma_1/\gamma_n$= 1\\
\hline
Ba$_{0.68}$K$_{0.32}$Fe$_{2}$As$_{2}$\footnotemark[1] &\onlinecite{popovich} &38.5 &$\sim$120 &50 &$\alpha_1$=1.1, $\gamma_1/\gamma_n$= 0.5\\
& & & & &$\alpha_2$=3.3, $\gamma_1/\gamma_n$= 0.5\\
\hline
KFe$_{2}$As$_{2}$\footnotemark[2] &\onlinecite{fukazawa} &3.5 &$\sim$21 &69.1 &$\alpha_1$=0.3, $\gamma_1/\gamma_n$= 0.55\\
& & & & &$\alpha_2$=2.4, $\gamma_2/\gamma_n$= 0.45\\
\hline
Ba(Fe$_{0.925}$Co$_{0.075}$)$_{2}$As$_{2}$\footnotemark[1] &\onlinecite{hardy,hardy1} &21.4 &30 &23.8 &$\alpha_1$=0.95, $\gamma_1/\gamma_n$= 0.33\\
& & & & &$\alpha_2$=2.2, $\gamma_2/\gamma_n$= 0.67\\
\hline
Ba(Fe$_{0.92}$Co$_{0.08}$)$_{2}$As$_{2}$\footnotemark[1] &\onlinecite{gofryk} &20 &$\sim$22 &18 &$\alpha_1$=0.957, $\gamma_1/\gamma_n$= 0.38\\
& & & & &$\alpha_2$=2.175, $\gamma_2/\gamma_n$= 0.62\\
\end{tabular}
\end{ruledtabular}
\footnotetext[1]{Single crystal} \footnotetext[2]{Polycrystal}
\end{table*}

The compound BNFA, on the other hand, shows an anomaly in $C/T$
around 29.4 K (Fig. 2), which is marked by the superconducting
transition. The temperature where this anomaly appears is
consistent with the sharp superconducting transition in
magnetization measurements (see Fig. 1). The jump in the specific
heat is reasonably pronounced with a $\Delta C/T_c$ value around
84 mJ mol$^{-1}$ K$^{-2}$ which is comparable to other hole
(potassium) doped 122 compounds with a quantity around 100 mJ
mol$^{-1}$ K$^{-2}$.\cite{mu122,popovich,welp} It
is worth to mention that the obtained $\Delta C/T_c$ for the
present material scales well with its $T_c$ in perspective of
recent results of Fe-pnictides.\cite{budko} The specific heat
measured in a magnetic field of 90 kOe (Fig. 2) shows that the
superconducting transition is broadened and insignificantly
shifted ($\sim$ 1.5 K) in 90 kOe for BNFA, which is expected from
its high $H_{c2}$ estimated to be above 100 Tesla.\cite{welp,weng}

In the inset of Fig. 2, specific heat data have been plotted in
the form $C/T$ vs $T^2$ for the compounds BFA and BNFA. At low
temperature, the data can be linearly fitted to $C/T$ = $\gamma$ +
$\beta T^2$, where $\gamma$ and $\beta$ are the electronic and
lattice coefficients of the specific heat.\cite{gopal} For BFA, we
obtain $\gamma$ = 6.13(8) mJ mol$^{-1}$ K$^{-2}$ and $\beta$ =
0.369(7) mJ mol$^{-1}$ K$^{-4}$. From the obtained $\beta$-value,
we calculate the Debye temperature $\theta_D$ following the
relation $\theta_D$ = [(12$\pi^4Rn$)/(5$\beta$)]$^{1/3}$, where
$R$ is the molar gas constant, and $n$ is the number of atoms per
formula unit.\cite{gopal} This gives $\theta_D$ = 297 K for BFA.
The extracted value of $\gamma$ in our crystal is consistent with
other studies on single crystals of BFA ($\sim$ 6.1 mJ
mol$^{-1}$ K$^{-2}$),\cite{dong,sefat} and close to
theoretically predicted values, i.e., 5.68 mJ mol$^{-1}$ K$^{-2}$
(Ref. \onlinecite{ma}) or 7.22 mJ mol$^{-1}$ K$^{-2}$ (Ref.
\onlinecite{singh}). The fact that the specific heat data $C/T$
versus $T^2$ for BNFA exhibit a linear behavior at low
temperatures without any upturn discards the possibility of
Schottky-like contributions in our sample under study.

\begin{figure}
    \centering
        \includegraphics[width=8cm]{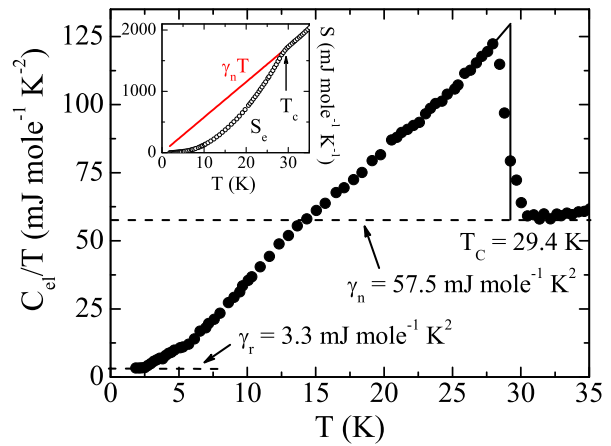}
    \caption{The electronic specific heat $C_{el}/T$ as function of temperature for the sample Ba$_{0.65}$Na$_{0.35}$Fe$_2$As$_2$. $\gamma_n$ and $\gamma_r$ represent the normal state and residual electronic coefficient of the specific heat. The inset shows the entropy in the normal and superconducting state as a function of temperature.}
    \label{fig:Fig3}
\end{figure}

The phononic contribution $C_{ph}$ to the specific heat of
BFA has been determined following the relation
$C_{ph}^{BFA}$ = $C_{tot}^{BFA}$ - $C_{el}^{BFA}$, where the
$C_{el}^{BFA}$ is $\gamma^{BFA}\cdot T$. We find
that the specific heat is dominated by phonons in this region,
i.e., around $T_c$, $C_{el}$ is only about 10\% of $C_{ph}$. Using
$C_{el}^{BNFA}/T$ = $C_{tot}^{BNFA}/T$ - $f \cdot C_{ph}^{BFA}/T$,
we can calculate $C_{el}^{BNFA}$. The scaling factor $f$ has been
introduced due to slightly different atomic compositions between
BNFA and BFA. To determine the value of $f$, we have used a criterion
that normal- and superconducting-state entropy are equal at $T_c$,
i.e., $\int_{0}^{T_c}\left(C_{el}/T\right)dT$ = $\gamma_nT_c$,
where $\gamma_n$ is the normal-state electronic specific heat
coefficient. We started with $f$ = 1 but the entropy
conservation criterion is satisfied for $f$ = 0.95 (inset of Fig. 3).
This practice yields $T_c$ = 29.4 K. The resulting $C_{el}/T$ for BNFA
is presented in the main panel of Fig. 3. It is obvious from the
figure that the superconducting transition at $T_c$ is reasonably
sharp, yielding a jump in $C_{el}/T$ at $T_c$ around 72.5 mJ
mol$^{-1}$ K$^{-2}$. From our determined $\gamma_n$ = 57.5 mJ
mol$^{-1}$ K$^{-2}$, we estimate the universal parameter
$C_{el}/\gamma_nT_c$ = 1.26. This value, however, is lower than
the weak-coupling BCS prediction of 1.43.\cite{poole} Following
the fact that the superconducting anomaly at $T_c$ is reasonably
sharp in BNFA, therefore a distribution in $T_c$ or the presence
of an impurity phase is an unlikely explanation for such a reduced
value of $C_{el}/\gamma_nT_c$. Instead, we believe that the
presence of multiple SC gaps possibly render a low
$C_{el}/\gamma_nT_c$ in BNFA, as evidenced in other 122
Fe-pnictides.\cite{hardy,gofryk,fukazawa} Moreover, the signature
of a multi-gap scenario in BNFA is evidenced by a significant hump
around 12 K in our $C_{el}/T$ vs $T$ data (Fig. 3), which will be
discussed below. Note that $C_e/T$ almost saturates at low
temperature, however, it does not extrapolate to zero, yielding a
residual electronic specific heat value $\gamma_r$ = 3.3 mJ
mol$^{-1}$ K$^{-2}$. We mention that the presence of a finite
$\gamma_r$ is common in both electron- and hole-doped 122
crystals,\cite{mu122,hardy,hardy1,kant,popovich} and that the
value of $\gamma_r$ in our present case is remarkably low, showing
the good quality of our single crystal. The origin of $\gamma_r$
in BNFA is not clear, however, it may arise due to an
incomplete transition to the superconducting state or broken
pairs in the superconducting condensate.\cite{hardy,hardy1,fisher}
Nonetheless, assuming a superconducting volume fraction
($\gamma_n$ - $\gamma_r$)/$\gamma_n$ $\approx$ 95\%, our crystal
consists of a high value, which is in fair agreement
with our magnetization data (Fig. 1).

The obtained high value of $\gamma_n$ for BNFA is consistent with
other members in the hole-doped 122 series whereas for
electron-doped 122 compounds $\gamma_n$ is much lower (see Table
I). Utilizing our value for $\gamma_n$, we can obtain information
about the normal state electronic properties, i.e., the DOS at the
Fermi energy $N(\epsilon_F)$ of BNFA using the
relation:\cite{kittel}

\begin{eqnarray}
    \gamma_n = \gamma_0\left(1 + \lambda\right),
\end{eqnarray}

\begin{eqnarray}
    \gamma_0 = \frac{\pi^2 k_B^2}{3}N(\epsilon_F),
\end{eqnarray}

where $\lambda$ is the electron-phonon coupling
constant and $k_B$ is the Boltzmann constant. Since in the case
of Fe-pnictides $\lambda$ is not significant, we can set
$\gamma_n$ $\equiv$ $\gamma_0$. Therefore, $\gamma_n$ is mainly
contributed by $N(\epsilon_F)$ which implies a higher
$N(\epsilon_F)$ in hole-doped compounds than in electron-doped
ones. From our $\gamma_n$, we calculate $N(\epsilon_F)$ = 24.37
states eV$^{-1}$ f.u.$^{-1}$. It is worth to mention that the
general high values of $\gamma_n$ or $N(\epsilon_F)$ for
hole-doped 122 Fe-pnictides remains controversial with theoretical
calculations yielding $\gamma_n$ = 13.03 mJ mol$^{-1}$ K$^{-2}$
and $N(\epsilon_F)$ = 5.526 states eV$^{-1}$ f.u.$^{-1}$ for
Ba$_{0.5}$K$_{0.5}$Fe$_2$As$_2$, where the related values are only
around 20\% higher than the parent compound.\cite{shein} However,
another calculation clarifies that upon including the band
parameters from experimental ARPES data as well as mass
renormalization effects, the calculated $\gamma_n$ is close to the
experimental values.\cite{benfatto} Nonetheless, this controversy
calls for further rigorous theoretical investigations adopting
possible reconciliations within the experimental findings.

\begin{figure}
    \centering
        \includegraphics[width=8cm]{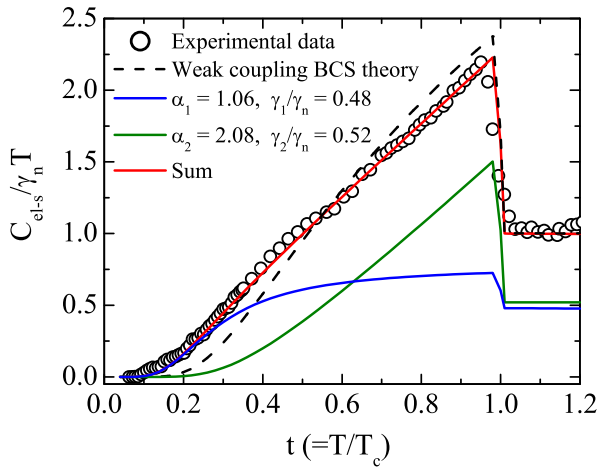}
    \caption{The normalized superconducting electronic specific heat ($C_{el-s}/\gamma_nT$) of Ba$_{0.65}$Na$_{0.35}$Fe$_2$As$_2$ as a function of reduced temperature $t = T/T_c$. The dashed line represents the theoretical curve based on single-band weak coupling BCS theory with the $s$-wave gap $\Delta(0)/k_BT_c$ = 1.76 following Eq. 4 and 5. The solid lines represent the curves of the two $s$-wave gap model (see text).}
    \label{fig:Fig4}
\end{figure}

After exploring the electronic specific heat and electronic
structure, we now examine the superconducting gap properties in
BNFA. In many cases, specific heat measurements have already been
proved to be an effective tool in understanding the
superconducting gap structure and its
distributions.\cite{padamsee,bouquet} For our sample, however,
$C_{el}$ first needs to be corrected due to a finite $\gamma_r$.
At low temperature $C_{el}$ is assumed to be contributed by the
superconducting ($C_{el-s}$) as well as the non-superconducting
normal ($C_{n}$) parts of the specific heat. While the normal
electrons will have specific heat contributions linear in
temperature ($\gamma_rT$) the superconducting electronic
contribution will be scaled by 1 - $\gamma_r/\gamma_n$. On this
basis, the sum of the individual contributions to $C_{el}$ allows
to extract $C_{el-s}$ from the following relation:

\begin{eqnarray}
    C_{el-s}/T = \frac{\gamma_n}{\gamma_n - \gamma_r}\left(C_{el}/T -
    \gamma_r\right).
\end{eqnarray}

In Fig. 4 we present the normalized data $C_{el-s}/\gamma_nT$ as a
function of reduced temperature $t (= T/T_c)$ for BNFA. As
mentioned earlier, $C_{el-s}/\gamma_nT$ exhibits a broad hump
around $t$ = 0.4, which implies the presence of multiple SC gaps
in this compound. We have analyzed our specific heat data
utilizing the $\alpha$-model which was originally proposed to
account for the thermodynamic properties of a strongly
coupled single-gap superconductor under semiempirical
approach.\cite{padamsee} This model, however, later had been
generalized to explain the specific heat behavior in multi-band,
multi-gap superconductors, i.e., MgB$_2$.\cite{bouquet} Following
this model, the thermodynamic properties like the entropy ($S$)
and $C$ can be calculated for a system of independent
quasiparticles as:\cite{padamsee,bouquet}

\begin{eqnarray}
    \frac{S}{\gamma_nT_c} = -\frac{6}{\pi^2}\frac{\Delta(0)}{k_BT_c} \int_0^\infty [f\ln f + (1 - f)\ln (1 - f)] dy,
\end{eqnarray}

\begin{eqnarray}
 \frac{C}{\gamma_nT_c} = t\frac{d\left(S/\gamma_nT_C\right)}{dt},
\end{eqnarray}

where $f$ = $\left[exp\left(\beta E\right) + 1\right]^{-1}$ and
$\beta$ = $(k_BT)^{-1}$. The energy of the quasiparticles is given
by $E$ = $\sqrt{\left[\epsilon^2 + \Delta^2(t)\right]}$, where
$\epsilon$ is the energy of the normal electrons relative to the
Fermi surface. In Eq. 4, the integration variable $y$ =
$\epsilon/\Delta(0)$, where $\Delta(0)$ is the zero temperature
gap magnitude and the scaled gap $\alpha$ = $\Delta(0)/k_BT_c$ is
the only adjustable fitting parameter. The temperature dependence
of the gap is determined as $\Delta(t)$ = $\Delta(0)\delta(t)$,
with $\delta(t)$ being obtained from the table in Ref.
\onlinecite{muhl}. In the case of two gaps, the thermodynamic
properties are determined as the sum of contributions from the two
gaps, i.e., $\alpha_1$ (= $\Delta_1(0)/k_BT_c$) and $\alpha_2$ (=
$\Delta_2(0)/k_BT_c$) with their respective weights
$\gamma_1/\gamma_n$ and $\gamma_2/\gamma_n$ respectively, where
$\gamma_1$ + $\gamma_2$ = $\gamma_n$.

Using Eqs. 4 and 5, we first calculate the specific heat
$C_{el-s}/\gamma_nT$ as a function of $t$ with $\alpha$ = 1.76 for
the single-band weak coupling BCS theory. As evident from Fig. 4,
the calculations disagree significantly with our experimental
data, where the former is characterized by a higher jump anomaly
at $T_c$. Moreover, an opposite curvature and different magnitude
below and above $t \approx$ 0.55 can be observed for the single-gap
model compared to the experimental data. We then calculate
$C_{el-s}/\gamma_nT$ introducing two gaps and their appropriate
weights. Apparently, values $\alpha_1$ = 1.06 [$\Delta_1$(0) =
2.68 meV], $\gamma_1/\gamma_n$ = 0.48 and $\alpha_2$ = 2.08
[$\Delta_2$(0) = 5.27 meV], $\gamma_2/\gamma_n$ = 0.52, yield the
closest matching with our experimental data (see Fig. 4). The Fig. 4
also shows the $C_{el-s}/\gamma_nT$ vs $t$ plot for an individual
$\alpha$ and its weights.

For the sake of comparison, we have summarized the values of
scaled gaps $\alpha_1$ and $\alpha_2$ and their respective
weights, $T_c$, $\Delta C_{el}/T_c$ and $\gamma_n$ for
Ba$_{0.65}$Na$_{0.35}$Fe$_2$As$_2$ along with other hole- and
electron-doped 122 materials in Table I. For BNFA, the larger gap
$\alpha_2$ has a higher value than the weak-coupling BCS gap value
(1.76) while the smaller one $\alpha_1$ has a lower value.
Although the gap magnitudes are scattered for different compounds
within the Ba-122 family, their relative weights exhibit a
consistent trend. Upon electron doping the smaller gap has around
(30 - 40)\% contribution to the electronic specific heat, whereas
for hole doped compounds both the bands contribute almost equally.
While our obtained gap structure for
Ba$_{0.65}$Na$_{0.35}$Fe$_2$As$_2$ has qualitative similarity with
other hole doped materials, such as
Ba$_{0.68}$K$_{0.32}$Fe$_2$As$_2$, the gap ratio
$\Delta_2$/$\Delta_1$ differs significantly (Table I).

In the scenario of an interband pairing model for
Fe-pnictides,\cite{dolgov1} the gap ratio is predicted as
$\Delta_2$/$\Delta_1$ = $\sqrt{N_1/N_2}$, where $N_1$ and $N_2$
are the Fermi-level DOS in the respective bands, and
$\Delta_2$/$\Delta_1$ is shown to evolve with the effective
coupling among the bands. Therefore, one can speculate that the
DOS in different bands as well as their coupling modify with
K$^{1+}$ and Na$^{1+}$ doping. Indeed, K$^{1+}$ and Na$^{1+}$ have
different ionic sizes and electronic configurations which may
contribute differently to these issues. However,
$\Delta_2$/$\Delta_1$ ratio implies that $N_1/N_2$ in
Ba$_{0.68}$K$_{0.32}$Fe$_2$As$_2$ is surprisingly about twice
higher than in Ba$_{0.65}$Na$_{0.35}$Fe$_2$As$_2$, which seems to
be an unlikely situation with just this kind of dopant variation
(see Table 1). On the other hand, the fact that $\gamma_1 \sim
\gamma_2$ $\sim$ $0.5\gamma_n$ for both compounds implies that
both gaps open up at the FS with almost equal DOS irrespective of
the dopant species. These observations probably suggest that the
theoretical discussions in Ref. \onlinecite{dolgov1} need to include
more than two bands. In fact, ARPES results\cite{ding,evtu1}
strongly hint towards the inclusion of at least four bands (two
hole-like and two electron-like) opening the superconducting gaps
in Ba$_{1-x}$K$_{x}$Fe$_2$As$_2$. Also, the need for four bands to
describe the thermodynamic signatures has been pointed out in
theoretical calculation.\cite{benfatto} At the same time, one can
clearly see in Table 1 that within multi-gap analysis the smaller
gap $\alpha_1$ remains almost close to 1 for all materials (except
for the extremely hole-doped case KFe$_2$As$_2$). However, the larger one
$\alpha_2$, which appears in the strongly nested inner hole-like
and electron-like bands, varies with both the doping element and
their concentration, illustrating that the nesting condition is
modified with the doping in Fe-pnictides, which is quite
intriguing.

It is worth to mention here that the employed $\alpha$-model follows a simple semiempirical approach where the superconducting gap is
assumed to have BCS temperature dependence and the interband coupling is not taken into account. Despite such simplification this model has been
extensively used to analyze the experimental thermodynamic data for many kinds of materials. One, however, certainly needs to check other
self-consistent models to compare the results. In this scenario, within the framework of Eliashberg approach for MgB$_2$ Dolgov \textit{et
al.}\cite{dolgov2} has shown that $\alpha$-model is sufficiently accurate to find the superconducting gap values though the gap's partial
contribution lacks in full agreement. Another recently proposed $\gamma$-model by Kogan \textit{et al.}\cite{kogan} is also an effective approach
which takes into account the interband pairing potential and is successfully tested for two band superconductors MgB$_2$ and V$_3$Si. Our
experimental work calls therefore for a detailed theoretical analysis of our data with these and other appropriate models to fully understand the
multigap superconducting nature in Fe-pnictides.
In addition, considering the fact that superconducting
gaps estimated by using different experimental techniques like
ARPES, PCARS, or muon spin rotation ($\mu$SR) exhibit a wide
distribution of absolute values,\cite{evtu2} further studies
involving specific heat measurements are required on doped 122
Fe-pictides with different kinds of doping elements as well as
doping concentrations to develop a comprehensive understanding and
a generalized view on this matter.

\section{Conclusions}
In summary, the electronic properties and superconducting gap
structure of hole doped 122 Fe-pnictide
Ba$_{0.65}$Na$_{0.35}$Fe$_2$As$_2$ are studied by measuring
specific heat. A reasonably pronounced anomaly has been found around
$T_c$ = 29.4 K. In applied magnetic fields,
the transition becomes broadened, however, there is only a minimal
decrease in $T_c$ of about 1.5 K in 90 kOe. Employing an entropy
conservation criterion at $T_c$, we extract $\gamma_n$ = 57.5 mJ
mole$^{-1}$ K$^{-2}$ which agrees well with other hole doped 122
compounds. It is interesting that the temperature-dependent
superconducting electronic specific heat cannot be explained
within single-band weak-coupling BCS theory. From our analysis we
find that the presence of $s$-wave like two gaps with magnitudes
$\Delta_1(0)/k_BT_c$ = 1.06 and $\Delta_2(0)/k_BT_c$ = 2.08 and
respective weights of about $\gamma_1$/$\gamma_n$ = 0.48 and
$\gamma_2$/$\gamma_n$ = 0.52 matches well with our experimental
data. Though these results are qualitatively similar to K-doped
122 compounds, on a quantitative level their gap magnitudes and
their ratios are quite different. This calls for further studies
on materials with different doping levels to reach a full
understanding of the gap structure and related mechanisms.

\section{Acknowledgment}
We acknowledge fruitful discussions with V. Zabolotnyy. We thank
M. Deutschmann, S. M\"uller-Litvanyi, R. M\"uller, J. Werner, and
S.~Ga{\ss} for technical support. This work has been supported by
the DFG, Germany through grant no BE 1749/13 and WO 1532/3-1.

\end{document}